\documentclass[prd,aps,twocolumn,superscriptaddress,showkeys,10pt]{revtex4-2}
\usepackage{amsmath,graphicx,amssymb}
\usepackage{multirow}
\usepackage{mathptmx} 
\usepackage{lipsum}
\usepackage{latexsym}
\usepackage{booktabs}
\usepackage{caption}
\usepackage{subcaption}
\RequirePackage{flushend}
\RequirePackage[colorlinks,citecolor=blue,urlcolor=blue,linkcolor=blue]{hyperref}
\usepackage{orcidlink}
\begin{document}

	\title{Geometrothermodynamics characterization of ABG-AdS black hole immersed by string clouds}
\author{Sayed Farooq Ahmed Shah \orcidlink{0000-0002-6778-2305}}
\email{sshah@gudgk.edu.pk}
\affiliation{College of Electrical and Mechanical Engineering, National University of Sciences and Technology, Rawalpindi, 46000, Pakistan}
\affiliation{Department of Mathematics, Ghazi University, Dera Ghazi Khan, 32200, Pakistan}

\author{Safia Taj \orcidlink{0009-0006-0714-1350}}
\email{safiataaj@ceme.nust.edu.pk}
\affiliation{College of Electrical and Mechanical Engineering, National University of Sciences and Technology, Rawalpindi, 46000, Pakistan}

\date{\today}

\begin{abstract}
Regular black holes, whose singular cores are replaced by smooth geometries sourced by nonlinear electrodynamics, provide a natural setting to study how thermodynamic stability is encoded in spacetime geometry. Conventional stability criteria, however, depend on the choice of thermodynamic potential. Here we develop the quasi-homogeneous geometrothermodynamics of the regular Ay\'on-Beato--Garc\'ia anti-de Sitter (ABG-AdS) black hole surrounded by a cloud of strings, a Legendre-invariant framework in which no potential needs to be preferred. We show that the entropy, electric monopole charge and AdS radius act as extensive variables obeying a generalized quasi-homogeneity condition, and that the equilibrium manifold then carries three families of thermodynamic metrics. We establish an exact correspondence between response-function anomalies and curvature singularities: the first metric's curvature diverges on the zero-temperature extremal boundary, where the heat capacity vanishes, signaling Davies first-order transitions, whereas the second's diverges
precisely where the heat capacity does, pinpointing Davies second-order transitions. The Gibbs free energy displays a swallow-tail structure that reveals a first-order small/large black hole transition, whose metastable region is bounded by these divergences of the heat capacity, together with a Hawking--Page-type transition to thermal AdS. The string-cloud density and the electric charge jointly control the horizon structure, the admissible thermal domain and the stability intervals.
\end{abstract}
\maketitle

\section{Introduction}
\label{sec:1}

Gravitational-wave ringdowns from coalescing compact binaries detected by the LIGO--Virgo--KAGRA collaboration~\cite{Abbott2016d,Abbott2020} and horizon-scale images of $\text{M87}^*$ and $\text{Sgr~A}^*$ captured by the Event Horizon Telescope~\cite{EventHorizonTelescope2019a,EventHorizonTelescope2019b,EventHorizonTelescope2019c,EventHorizonTelescope2019d} have firmly established black holes as direct probes of strong-field gravity. Complementing these observational milestones, the foundational works of Bekenstein, Bardeen, Carter, Hawking, and Smarr \cite{Bekenstein1973,Hawking1975,BardeenCarterHawking1973,Smarr1973} demonstrated that black holes are genuine thermodynamic systems. General relativity (GR), quantum field theory, and statistical mechanics are all connected by the laws of black hole mechanics that directly mirror those of classical thermodynamics when the surface gravity of the horizon is identified with the Hawking temperature and its area with the Bekenstein--Hawking entropy. This interplay becomes especially rich in anti-de Sitter (AdS) spacetimes, where the negative cosmological constant admits a gauge/gravity dual description~\cite{Witten1998} and gives rise to diverse phase transitions, ranging from the Hawking--Page transition~\cite{Hawking1983} to van der Waals--type critical behavior~\cite{Chamblin1999,Kubiznak2012}.

Classical GR, however, predicts its own breakdown. The Penrose--Hawking singularity theorems~\cite{Penrose1965,HawkingPenrose1970} establish that gravitational collapse generically terminates in curvature singularities, where predictability is lost. Building on the proposals of Sakharov~\cite{Sakharov1966} and Gliner~\cite{Gliner1966} that matter at extreme densities may settle into a de~Sitter core, Bardeen constructed the first black hole metric free of curvature singularities~\cite{Bardeen1968}. Subsequent regular geometries, notably those of Dymnikova~\cite{Dymnikova1992} and Hayward~\cite{Hayward2006}, have since been investigated through quasinormal modes, geodesic motion, wormhole stability, and strong lensing. A decisive step was taken by Ay\'on-Beato and Garc\'ia (ABG), who showed that regular black holes arise as exact solutions of GR minimally coupled to a nonlinear electrodynamics (NED) source~\cite{AyonBeatoGarcia1998,AyonBeatoGarcia2000,AyonBeatoGarcia2005}. The AdS generalization of these solutions thus provides a controlled setting in which to study how nonlinear electromagnetic self-interaction reshapes horizon thermodynamics and critical behavior~\cite{Singh2021}.

Realistic black holes, moreover, are not isolated but embedded in extended matter and relic-field distributions that deform the background geometry. Motivated by string theory, in which one-dimensional strings replace point particles as elementary constituents, Letelier~\cite{letelier1979clouds} proposed a cloud of strings as an anisotropic, pressureless fluid source for the Einstein field equations. Characterized by a density parameter $a$, this cloud modifies both the near-horizon and asymptotic structure of the spacetime. For the ABG-AdS black hole immersed by string clouds, the interplay between the NED electric monopole charge and the string background has been shown to affect test-particle chaos, greybody spectra, and shadow observables \cite{Kumar2024,Liu2026}. The thermodynamic stability and phase structure of this system, by contrast, have yet to be analyzed within a framework independent of the choice of thermodynamic potential.

Geometrothermodynamics (GTD), introduced by Quevedo~\cite{Quevedo2007,Quevedo2008}, provides precisely such a framework. It combines the contact geometry of the $(2n+1)$-dimensional thermodynamic phase space $\mathcal{T}$ with the Riemannian geometry of the $n$-dimensional equilibrium manifold $\mathcal{E}$. By treating Legendre transformations as coordinate changes on $\mathcal{T}$, the phase-space metrics are constructed to preserve their functional form under these mappings. Consequently, the metrics induced on $\mathcal{E}$ via pullback remain entirely invariant regardless of the thermodynamic potential chosen to describe the system. This formalism has also been extended to quasi-homogeneous systems, whose fundamental equations satisfy generalized scaling symmetries with distinct quasi-homogeneity coefficients, allowing black holes with multiple extensive scales to be treated consistently~\cite{Quevedo2019,Quevedo2023,Shah2026,Quevedo2026}. Within this setting, a nonvanishing scalar curvature indicates the presence of thermodynamic interactions, whereas scalar curvature singularities mark the breakdown of equilibrium thermodynamics and the onset of phase transitions.

We construct a quasi-homogeneous GTD framework for the regular ABG-AdS black hole with a string cloud. In the mass representation, the fundamental relation exhibits generalized quasi-homogeneity when the Bekenstein–Hawking entropy \(S\), electric charge \(Q\), and AdS radius \(l\) are treated as extensive variables. Next, within the resulting Legendre-invariant framework, we construct three families of thermodynamic metrics on the equilibrium manifold and analytically derive the Hawking temperature, the conjugate potentials, and the heat capacity. Later, we establish an exact correspondence between response-function anomalies and scalar curvature singularities. The singularities of the first metric, $g^{\mathrm{I}}$, lie on the zero-temperature extremal boundary, where the heat capacity vanishes, and characterize first-order phase transitions (FOPT). Those of the second metric, $g^{\mathrm{II}}$, coincide with the divergences of the heat capacity and characterize Davies-type second-order phase transitions (SOPT). Furthermore, the Gibbs free energy displays a swallow-tail structure that reveals a FOPT between small and large black holes, whose metastable region is bounded by these divergences of the heat capacity, together with a Hawking--Page-type transition to thermal AdS. Finally, we show how the interplay between string-cloud density $a$ and the electric charge $Q$ governs the horizon structure, the physically admissible thermal domain, and the stability intervals.

The remainder of this paper is organized as follows. In Sec.~\ref{sec:2}, we introduce the action for Einstein gravity coupled to NED and a cloud of strings, and we present the corresponding static, spherically symmetric ABG-AdS solution. We then examine its horizon structure, quasi-homogeneity relations, and classical thermodynamic stability in the mass representation. In Sec.~\ref{sec:3}, we review the GTD formalism for quasi-homogeneous systems, construct three families of thermodynamic metrics on the equilibrium manifold, and calculate the corresponding scalar curvatures. In Sec.~\ref{sec:4}, we study the heat capacity, the Gibbs free energy, and the trace of its Hessian matrix, and we discuss the resulting FOPT and SOPT by comparing them with the curvature singularities obtained in Sec.~\ref{sec:3}. Finally, Sec.~\ref{sec:5} presents our concluding remarks.
\section{The ABG-AdS Black Hole with string clouds}\label{sec:2}

We investigate the regular ABG-AdS black hole immersed by string clouds within Einstein gravity. The gravitational system involves a NED source and a string cloud characterized by the density parameter \(a\). Section~\ref{sec:2.1} presents the gravitational action, spacetime geometry, and horizon structure, while Sec.~\ref{sec:2.2} develops the thermodynamic formulation, including the first law and quasi-homogeneous Smarr relation.

\subsection{The metric}\label{sec:2.1}

We construct static, spherically symmetric solutions in which the ABG spacetime is supplemented by a negative cosmological constant and a surrounding cloud of strings. The system consists of Einstein gravity coupled to a NED field and a cloud of strings, with the action
\begin{equation}
	I = \int d^{4}x \sqrt{-g}\left(R + 2\Lambda + \mathcal{L}(F)\right) + I_{\rm cs}.
	\label{e:1}
\end{equation}
Here $R$ is the Ricci scalar, $g=\det g_{\mu\nu}$, $\Lambda<0$ corresponds to an AdS background, and $\mathcal{L}(F)$ is the NED Lagrangian, a function of $F=\tfrac14 F_{\mu\nu}F^{\mu\nu}$ with $F_{\mu\nu}=\partial_\mu A_\nu-\partial_\nu A_\mu$. The strings are described by the Nambu--Goto action $I_{\rm cs}=\int \sqrt{-\gamma}\,\mathcal{M}\,d\xi^0 d\xi^1$, where $\mathcal{M}$ is a dimensionless constant, $\xi^0$ and $\xi^1$ are the timelike and spacelike worldsheet parameters, and $\gamma$ is the determinant of the induced metric $\gamma_{AB}=g_{\mu\nu}\,\partial_A x^{\mu}\partial_B x^{\nu}$.

The NED sector is conveniently treated through the auxiliary field $P_{\mu\nu}=\mathcal{L}_F F_{\mu\nu}$, with $\mathcal{L}_F=\partial\mathcal{L}/\partial F$ and $P=P_{\mu\nu}P^{\mu\nu}$. The Legendre transform $\mathcal{H}=2F\mathcal{L}_F-\mathcal{L}$ gives $\mathcal{L}=2P\mathcal{H}_P-\mathcal{H}$, with $\mathcal{H}_P=\partial\mathcal{H}/\partial P$ and $F_{\mu\nu}=\mathcal{H}_P P_{\mu\nu}$, so the theory is fixed by the structure function $\mathcal{H}(P)$. Varying the action \eqref{e:1} with respect to the metric and the gauge potential yields
\begin{equation}
	R_{\mu\nu} - \tfrac12 g_{\mu\nu}R + \Lambda g_{\mu\nu} = 8\pi\left( T_{\mu\nu}^{\rm EM} + T_{\mu\nu}^{\rm cs}\right),
	\label{e:2}
\end{equation}
\begin{equation}
	\nabla_\mu P^{\mu\nu}=0,
	\label{e:3}
\end{equation}
where the energy--momentum tensors of the NED field and of the string cloud are
\begin{equation}
	T_{\mu\nu}^{\rm EM} = \frac{1}{4\pi}\left[\mathcal{H}_P\, P_{\mu\lambda}P_{\nu}^{\ \lambda} - g_{\mu\nu}\left(2P\mathcal{H}_P-\mathcal{H}\right)\right],
	\label{e:4}
\end{equation}
\begin{equation}
	T_{\mu\nu}^{\rm cs} = \frac{\sigma}{8\pi}\, \frac{\Sigma_{\mu\alpha}\Sigma_{\nu}^{\ \alpha}}{\sqrt{-\gamma}},
	\label{e:5}
\end{equation}
and $\sigma$ is the string density.

For a purely electric configuration with charge $Q$, we take $P_{\mu\nu}=2\delta^{\theta}_{[\mu}\delta^{\phi}_{\nu]}D(r)$, so that $P=-Q^{2}/(2r^{4})$. The ABG black hole is sourced by
\begin{equation}
	\mathcal{H}(P)=\frac{P\left(1-3\sqrt{-2Q^{2}P}\right)}{\left(1+\sqrt{-2Q^{2}P}\right)^{3}}
	-\frac{3}{2Q^{2}s}\,\frac{(-2Q^{2}P)^{5/4}}{\left(1+\sqrt{-2Q^{2}P}\right)^{3/2}},
	\label{e:6}
\end{equation}
where $s=Q/2M$, and $Q$ and $M$ are free parameters related to the electric charge and the mass.
 This function reduces to Maxwell's theory, $\mathcal{H}(P)\rightarrow P$, for $P\ll1$, and satisfies the weak energy condition, $\mathcal{H}<0$ and $\mathcal{H}_P>0$. For the cloud of strings, the energy--momentum tensor takes the form
\begin{equation}
	T_{\mu\nu}^{\rm cs} = \frac{a}{8\pi r^{2}}\,\mathrm{diag}(1,1,0,0), \qquad 0<a<1,
	\label{e:7}
\end{equation}
with $a$ the string cloud parameter.

We adopt the static, spherically symmetric line element
\begin{equation}
	ds^2 = f(r)dt^2 - \frac{dr^2}{f(r)} - r^2\left(d\theta^2 + \sin^2\theta\, d\phi^2\right).
	\label{e:8}
\end{equation}
Substituting it into Eq.~\eqref{e:2} and using $\Lambda=-3/l^{2}$, with $l$ the AdS radius, the metric function of the ABG--AdS black hole with a string clouds is
\begin{equation}
	f(r) = 1 - a - \frac{2M r^{2}}{(r^{2}+Q^{2})^{3/2}} + \frac{Q^{2} r^{2}}{(r^{2}+Q^{2})^{2}} + \frac{r^{2}}{l^{2}}.
	\label{e:9}
\end{equation}
For $a=0$ the string cloud is absent, and Eq.~\eqref{e:9} recovers the standard ABG--AdS black hole.
The horizons of the black hole follow from the condition $f(r_h)=0$,
which can be solved for the mass parameter to give
\begin{equation}\label{e:10}
	M(r_h)=\frac{\left(Q^{2}+r_h^{2}\right)^{3/2}}{2r_h^{2}}
	\left[\,1-a+\frac{Q^{2}r_h^{2}}{\left(Q^{2}+r_h^{2}\right)^{2}}
	+\frac{r_h^{2}}{l^{2}}\right].
\end{equation}
Since $M(r_h)$ diverges both as $r_h\to0$ and as $r_h\to\infty$, it
possesses a single minimum, which corresponds to the extremal
configuration and is determined by
\begin{equation}
	f(r_e)=0,\qquad f'(r_e)=0 .
\end{equation}

\begin{figure}[htbp]
	\centering
	\includegraphics[width=\linewidth]{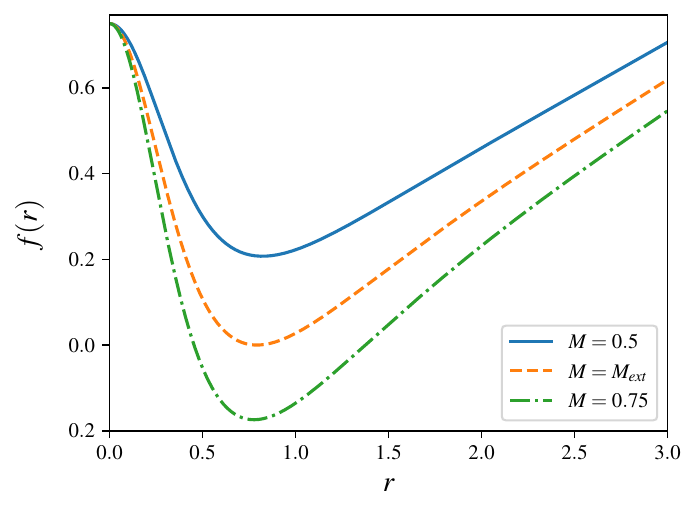}
\caption{Behavior of the metric function $f(r)$ versus radial coordinate $r$ for different $M$, with $Q=0.5$, $a=0.25$, and $l=6$.}
	\label{f:1}
\end{figure}

To illustrate the horizon structure, we choose the representative
parameter values
\begin{equation}
	Q=0.5,\qquad a=0.25,\qquad l=6 .
\end{equation}
For these values the extremal horizon is located at $r_e\simeq0.792$,
and the corresponding critical mass is $M_e\simeq0.6361$.
Figure~\ref{f:1} displays the metric function $f(r)$ for three
representative values of the mass parameter. For $M=0.5<M_e$, the
function stays positive for all $r$, so no horizon exists and the
solution describes a horizonless configuration. At $M=M_e$ the curve
just touches the horizontal axis at a single point, which signals a
degenerate horizon, i.e., an extremal black hole. For $M=0.75>M_e$ the
curve crosses the axis twice, giving two positive roots,
\begin{equation}
	r_-\simeq0.454,\qquad r_+\simeq1.375,
\end{equation}
which correspond to the inner and the outer horizons, respectively.
Hence, the critical mass $M_e$ separates the horizonless
configurations ($M<M_e$) from the nonextremal black holes ($M>M_e$).
\subsection{Thermodynamics}\label{sec:2.2}

In keeping with the Bekenstein--Hawking law, the entropy of a black hole (BH) equals one quarter of the area $A$ of its event horizon, $S=A/4$, and this relation establishes BH thermodynamics. For a spherically symmetric BH with horizon radius $r_h$, determined by $f(r_h)=0$ with $f(r)$ given in Eq.~\eqref{e:9}, the entropy becomes $S=\pi r_h^2$. Solving the horizon condition for the mass and eliminating $r_h$ in favor of $S$ yields
\begin{equation}
	\begin{split}
		M =\frac{\left(S -\pi  \left(a -1\right) l^{2}\right) \left(\pi  \,Q^{2}+S \right)^{2}+\pi^{2} Q^{2} l^{2} S}{2 \pi^{\frac{3}{2}} l^{2} S \sqrt{\pi  \,Q^{2}+S}},
	\end{split}
	\label{eq:1}
\end{equation}
which serves as the fundamental equation, $M=M(S,Q,l)$, of the ABG--AdS BH surrounded by a cloud of strings.

To clarify the physical meaning of the curvature singularities, we start from the first law of BH thermodynamics, which in the mass representation associated with Eq.~\eqref{eq:1} reads
\begin{equation}
	dM=T\,dS+\phi\,dQ+\Psi_{l}\,dl .
\end{equation}
The intensive variables appearing here are the Hawking temperature $T$, the potential $\phi$ conjugate to $Q$, and the variable $\Psi_{l}$ conjugate to $l$. Differentiating the mass function \eqref{e:10} gives
\begin{align}
	T &= \frac{\partial M}{\partial S}
	= \frac{\left[3S^{2}+\pi(a-1)l^{2}\left(2\pi Q^{2}-S\right)\right]\sqrt{S+\pi Q^{2}}}{4\pi^{3/2}l^{2}S^{2}}\nonumber \\
	&\quad - \frac{\sqrt{\pi}\,Q^{2}}{4\left(S+\pi Q^{2}\right)^{3/2}},\label{eq:2} \\
	\phi &= \frac{\partial M}{\partial Q}
	= \frac{3Q\left[S-\pi(a-1)l^{2}\right]\sqrt{S+\pi Q^{2}}}{2\sqrt{\pi}\,l^{2}S} \nonumber\\
	&\quad + \frac{\sqrt{\pi}\,Q\left(2S+\pi Q^{2}\right)}{2\left(S+\pi Q^{2}\right)^{3/2}}, \\
	\Psi_{l} &= \frac{\partial M}{\partial l}
	= -\frac{\left(S+\pi Q^{2}\right)^{3/2}}{\pi^{3/2}l^{3}} .
\end{align}

\begin{figure*}[htbp]
	\centering
	\begin{subfigure}[b]{0.48\textwidth}
		\centering
		\includegraphics[width=\textwidth]{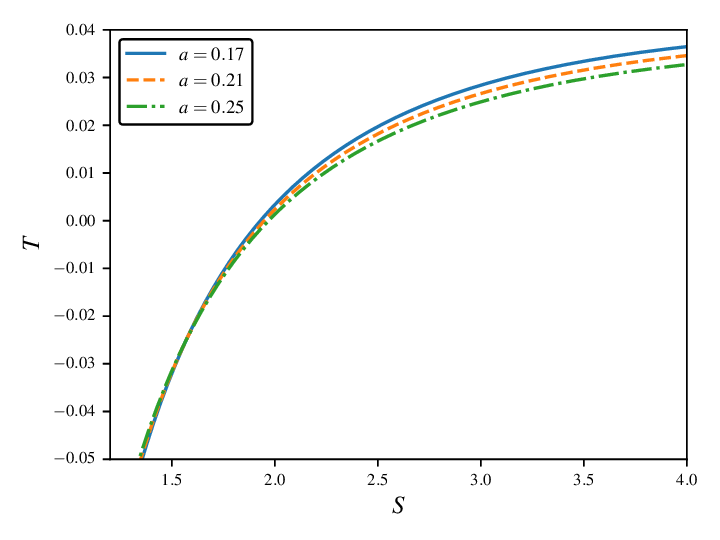}
		\caption{}
		\label{f:2a}
	\end{subfigure}
	\hfill
	\begin{subfigure}[b]{0.48\textwidth}
		\centering
		\includegraphics[width=\textwidth]{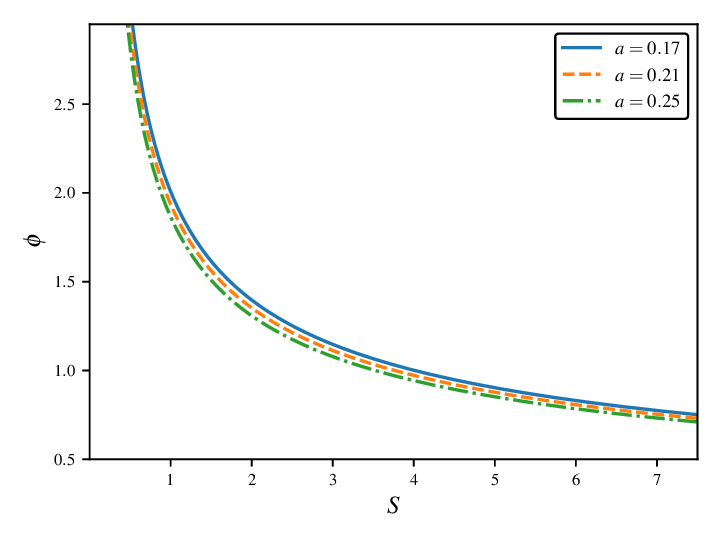}
		\caption{}
		\label{f:2b}
	\end{subfigure}
	\caption{Thermodynamic behavior of (a) Hawking temperature $T$ and (b) electric potential $\phi$ versus entropy $S$ for different string cloud parameter $a$, with fixed $Q=0.5$ and $l=6$.}
	\label{f:2}
\end{figure*}

Figure~\ref{f:2a} shows that the Hawking temperature $T$ rises monotonically with the entropy $S$ from negative to positive values, while a larger $a$ lowers $T$ and shifts its zero crossing to a higher $S$. Conversely, Fig.~\ref{f:2b} shows that the electric potential $\phi$ decreases monotonically with $S$, and that increasing $a$ uniformly reduces $\phi$.

In BH thermodynamics, fundamental equations are usually quasi-homogeneous rather than homogeneous. Accordingly, Eq.~\eqref{eq:1} must satisfy
\begin{equation}
	M(\lambda^{\beta_a}E^a)=\lambda^{\beta_M}M(E^a),
\end{equation}
where the $\beta_a$ are real constants called quasi-homogeneity coefficients and the variables $E^a$ are rescaled by a constant $\lambda>0$. Once the AdS radius $l$ is treated as a thermodynamic variable, Eq.~\eqref{eq:1} is only quasi-homogeneous, namely
\begin{equation}
	M(\lambda^{\beta_S}S,\lambda^{\beta_Q}Q,\lambda^{\beta_l}l)=\lambda^{\beta_M}M(S,Q,l),
\end{equation}
provided that the coefficients obey
\begin{equation}
	\beta_S=2\beta_Q,\quad \beta_Q=\beta_l,\quad \beta_S=2\beta_M .
\end{equation}

Quasi-homogeneity leaves the fundamental principles of thermodynamics intact, but it modifies standard relations such as the Gibbs--Duhem identity,
\begin{equation}
	\begin{split}
		&\sum_a(\beta_a-\beta_M)\frac{\partial M}{\partial E^a}\,dE^a \\
		&\quad +\sum_{a,b}\beta_aE^a\frac{\partial^{2}M}{\partial E^a\partial E^b}\,dE^b=0 .
	\end{split}
\end{equation}
The homogeneous case is recovered for $\beta_a=\beta_M=1$. In the mass representation, the Euler identity for the ABG--AdS BH surrounded by a cloud of strings is therefore
\begin{equation}
	2S\frac{\partial M}{\partial S}+Q\frac{\partial M}{\partial Q}+l\frac{\partial M}{\partial l}=M .
\end{equation}

\section{Geometrothermodynamics}
\label{sec:3}
Geometrothermodynamics recasts equilibrium thermodynamics in the language of differential geometry. Its guiding principle is that the physical content of a system should not depend on which thermodynamic potential is used to describe it. Since a Legendre transformation only exchanges a potential and its variables for another equivalent set, GTD regards it as a change of coordinates on a manifold whose geometry remains intact \cite{Quevedo2011}. To realize this principle, the manifold is endowed with a Riemannian metric that is itself Legendre invariant, and the resulting construction defines the thermodynamic phase space $\mathcal{T}$ \cite{Quevedo2023}. The equilibrium states of the system are then confined to an $n$-dimensional sub-manifold $\mathcal{E}\subset\mathcal{T}$, called the equilibrium space.

Consider a system with $n$ thermodynamic degrees of freedom. Its phase space $\mathcal{T}$ has dimension $2n+1$ and is coordinatized by $Z^A=\{\Phi, E^a, I_a\}$, where $\Phi$ is the thermodynamic potential, $E^a$ are the extensive variables, and $I_a$ are their conjugate intensive variables. A metric $G$ on $\mathcal{T}$ is Legendre invariant when its functional form is unchanged under such coordinate changes. Three families of metrics with this property are given by
\begin{align}
	G^{\mathrm{I}} &= (\mathrm{d}\Phi - I_a \mathrm{d}E^a)^2 
	+ (\zeta_{ab} E^a I^b)(\delta_{cd} \mathrm{d}E^c \mathrm{d}I^d), 
	\label{e:11} \\
	G^{\mathrm{II}} &= (\mathrm{d}\Phi - I_a \mathrm{d}E^a)^2 
	+ (\zeta_{ab} E^a I^b)(\eta_{cd} \mathrm{d}E^c \mathrm{d}I^d), 
	\label{e:12} \\
	G^{\mathrm{III}} &= (\mathrm{d}\Phi - I_a \mathrm{d}E^a)^2 
	+ \sum_{a=1}^{n} \zeta_a (E^a I^a)^{2k+1} \mathrm{d}E^a \mathrm{d}I^a,
	\label{e:13}
\end{align}
with $\delta_{ab}=\mathrm{diag}(1,\dots,1)$ and $\eta_{ab}=\mathrm{diag}(-1,1,\dots,1)$, while $\zeta_a$ and $\zeta_{ab}$ are real constants and $k$ is an integer.

The equilibrium space is introduced through the map $\varphi:\mathcal{E}\rightarrow\mathcal{T}$, with $E^a$ serving as coordinates on $\mathcal{E}$. This map is required to annihilate the contact form, that is,
\begin{equation} \label{e:14}
	\mathrm{d}\Phi - I_a \mathrm{d}E^a = 0.
\end{equation}
Because $\Phi=\Phi(E^a)$ is the fundamental equation of the system, the embedding is completely fixed by expressing every phase-space coordinate in terms of $E^a$:
\[
Z^A \rightarrow Z^A(E^a) = \{ \Phi(E^a), E^a, I_a(E^a) \}.
\]
The pullback $\varphi^*(G)=g=g_{ab}\,\mathrm{d}E^a\mathrm{d}E^b$ then induces a Riemannian metric on $\mathcal{E}$. Substituting Eqs.~\eqref{e:11}--\eqref{e:13} into this pullback yields the following metrics, which are the final output of the GTD construction:
\begin{align}
	g^{\mathrm{I}} &= \sum_{a,b,c=1}^{n} \left(\beta_c E^c \frac{\partial \Phi}{\partial E^c} \right) 
	\frac{\partial^2 \Phi}{\partial E^a \partial E^b} \mathrm{d}E^a \mathrm{d}E^b,
	\label{e:15} \\
	g^{\mathrm{II}} &= \sum_{a,b,c,d=1}^{n} \left(\beta_c E^c \frac{\partial \Phi}{\partial E^c} \right) 
	\eta^d_a \frac{\partial^2 \Phi}{\partial E^b \partial E^d} \mathrm{d}E^a \mathrm{d}E^b,
	\label{e:16} \\
	g^{\mathrm{III}} &= \sum_{a,b=1}^{n} \left(\beta_a E^a \frac{\partial \Phi}{\partial E^a} \right) 
	\frac{\partial^2 \Phi}{\partial E^a \partial E^b} \mathrm{d}E^a \mathrm{d}E^b,
	\label{e:17}
\end{align}
where $\eta^a_c=\mathrm{diag}(-1,1,\dots,1)$ and $\beta_a$ are real constants.

Consequently, a GTD analysis of a specific system can only be carried out once its fundamental equation $\Phi=\Phi(E^a)$ is known explicitly, since the thermodynamic properties and the geometry of $\mathcal{E}$ both follow from it. To evaluate the three invariant metrics in Eqs.~\eqref{e:15}--\eqref{e:17} for the present BH, we take the mass $M$ as the thermodynamic potential $\Phi$ and $E^a=(S,Q,l)$ as the extensive variables, which gives
\begin{align}
	g^{\mathrm{I}} &= \left(\beta_M M\right)\Big[M_{,SS}\,dS^2 + M_{,QQ}\,dQ^2 + M_{,ll}\,dl^2 \nonumber\\
	&\quad+ 2\big(M_{,SQ}\,dS\,dQ + M_{,Sl}\,dS\,dl + M_{,Ql}\,dQ\,dl\big)\Big], \label{e:18}\\
	g^{\mathrm{II}} &= \left(\beta_M M\right)\Big[-M_{,SS}\,dS^2 + M_{,QQ}\,dQ^2 + M_{,ll}\,dl^2 \nonumber\\
	&\quad+ 2 M_{,Ql}\,dQ\,dl\Big], \label{e:19}\\
	g^{\mathrm{III}} &= \beta_S\,S\,T\,M_{,SS}\,dS^2 + \beta_Q\,Q\,\phi\,M_{,QQ}\,dQ^2 + \beta_l\,l\,\Psi_l\,M_{,ll}\,dl^2 \nonumber\\
	&\quad+ \left(\beta_S\,S\,T + \beta_Q\,Q\,\phi\right)M_{,SQ}\,dS\,dQ \nonumber\\
	&\quad+ \left(\beta_S\,S\,T + \beta_l\,l\,\Psi_l\right)M_{,Sl}\,dS\,dl \nonumber\\
	&\quad+ \left(\beta_Q\,Q\,\phi + \beta_l\,l\,\Psi_l\right)M_{,Ql}\,dQ\,dl. \label{e:20}
\end{align}

As reported in \cite{Ladino2024}, three distinct classes of singularities have been identified. The corresponding conditions for the occurrence of Ricci scalar singularities in the GTD metrics defined by Eqs. \eqref{e:18}--\eqref{e:20} are expressed as follows:

\begin{align} 
	I&:  
3S^{4}+\pi\left[6Q^{2}+(1-a)l^{2}\right]S^{3} +\pi^{2}Q^{2}\left(3Q^{2}-l^{2}\right)S^{2} \nonumber\\
& -3\pi^{3}(1-a)l^{2}Q^{4}S -2\pi^{4}(1-a)l^{2}Q^{6}= 0, \label{e:21}\\
	II&:  3S^{5}+\pi\left[6Q^{2}-(1-a)l^{2}\right]S^{4} +8\pi^{5}(1-a)l^{2}Q^{8}\nonumber \\
	&+15\pi^{3}(1-a)l^{2}Q^{4}S^{2} +20\pi^{4}(1-a)l^{2}Q^{6}S \label{e:22}\\
	&+\pi^{2}Q^{2}\left[3Q^{2}+(5-2a)l^{2}\right]S^{3}+8\pi^{5}(1-a)l^{2}Q^{8}  =0, \nonumber\\
	III&: \sqrt{\pi  \,Q^{2}+S}=0 \label{e:23}
\end{align}
\section{Thermodynamic Stability and Phase Transitions} \label{sec:4}

Following the logic of ordinary equilibrium thermodynamics, Davies \cite{davies1977} proposed that the phase transitions of a BH can be located through the divergences of its heat capacity. Such transitions are usually interpreted as a change between small and large BHs, or equivalently between cold and hot ones. Since the microscopic degrees of freedom of a BH are still not understood, alternative tools can help us to understand these transitions better. Here we analyze the phase structure of the ABG--AdS BH with a cloud of strings through the heat capacity, the Gibbs free energy, and the trace of its Hessian matrix, and we compare the resulting transition points with the singularities of the Quevedo curvature scalars obtained in Sec.~\ref{sec:3}.

The local stability is governed by the heat capacity at fixed $a$, $l$, and $Q$,
\begin{equation}
	C = T\left(\frac{\partial S}{\partial T}\right)
	= \frac{N_1 N_2}{D_1}, \label{e:25}
\end{equation}
where
\begin{align}
	N_1 &=2S\left(S+\pi Q^{2}\right),\nonumber\\
	N_2 &=3S^{4}+\pi\left[6Q^{2}+(1-a)l^{2}\right]S^{3} +\pi^{2}Q^{2}\left(3Q^{2}-l^{2}\right)S^{2} \nonumber\\
	&\quad-3\pi^{3}(1-a)l^{2}Q^{4}S-2\pi^{4}(1-a)l^{2}Q^{6}, \nonumber\\
	D_1 &= 3S^{5}+\pi\left[6Q^{2}-(1-a)l^{2}\right]S^{4} \nonumber\\
	&\quad+\pi^{2}Q^{2}\left[3Q^{2}+(5-2a)l^{2}\right]S^{3} +15\pi^{3}(1-a)l^{2}Q^{4}S^{2} \nonumber\\
	&\quad+20\pi^{4}(1-a)l^{2}Q^{6}S+8\pi^{5}(1-a)l^{2}Q^{8}. \nonumber
\end{align}
Comparing with Eqs.~\eqref{e:21}--\eqref{e:23}, $N_2=0$ is condition $I$, $D_1=0$ is condition $II$, and condition $III$ corresponds to the factor $S+\pi Q^{2}$ of $N_1$, which vanishes only at the non-physical value $S=-\pi Q^{2}$. Unless stated otherwise, we take $Q=0.5$, $a=0.25$, and $l=6$. For these values, $C$ vanishes at $S=1.9688$ and diverges at the two positive roots of $D_1$, $S=9.5986$ and $S=18.8026$. The sign of $C$ splits the physical region into three branches: a stable small black hole (SBH) branch for $1.9688<S<9.5986$, an unstable intermediate black hole (IBH) branch for $9.5986<S<18.8026$ with $C<0$, and a stable large black hole (LBH) branch for $S>18.8026$.

\begin{figure}[ht]
	\centering
	\includegraphics[width=\columnwidth]{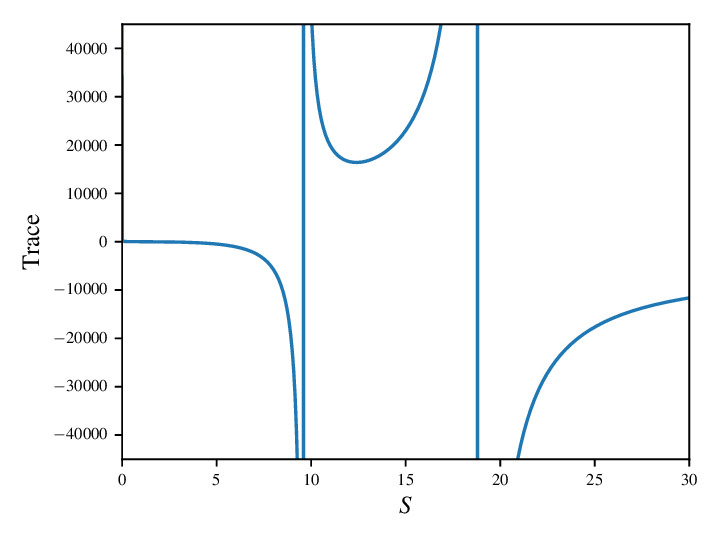}
	\caption{Behavior of the trace $\tau$ against entropy $S$ for $Q=0.5$, $a=0.25$, and $l=6$, displaying the critical divergences.}
	\label{f:3}
\end{figure}

To maintain global stability, we use the Gibbs free energy $G=H_M-TS$, where $H_M$ is the enthalpy, identified with the mass of the ABG--AdS spacetime. Using Eqs.~\eqref{eq:1} and \eqref{eq:2}, we obtain
\begin{equation}
	G = \dfrac{N_3}{D_3},\label{e:26}
\end{equation}
with
\begin{align}
	N_3 &= S(\pi Q^{2}+S)^{2}(2\pi Q^{2}-S)- a\pi l^{2}(\pi Q^{2}+S)^{2}(4\pi Q^{2}+S) \nonumber\\
	&\quad + \pi l^{2}\bigl(S^{3}+9\pi Q^{2}S^{2}+11\pi^{2}Q^{4}S+4\pi^{3}Q^{6}\bigr),\nonumber\\
	D_3 &= 4\pi^{3/2}\, S\, l^{2}\, (\pi Q^{2}+S)^{3/2}.\nonumber
\end{align}
The stability can also be assessed through the Hessian matrix $\mathbb{H}_{ij}=\partial^{2}G/\partial x_i\partial x_j$ with $x_i=(T,Q,l)$, and its trace
\begin{equation}
	\tau=\mathrm{Tr}(\mathbb{H})=\frac{\partial^2 G}{\partial T^2}+\frac{\partial^2 G}{\partial Q^2}+\frac{\partial^2 G}{\partial l^2},\label{eq:5}
\end{equation}
whose denominator is
\begin{align}
	\mathrm{denom}(\tau) &=
	8\pi^{5}(a-1)l^{2}Q^{8}+20\pi^{4}(a-1)l^{2}S\,Q^{6} -3S^{5}\nonumber\\ &\quad+15\pi^{3}(a-1)l^{2}S^{2}Q^{4}
	-3\pi^{2}S^{3}Q^{4}-6\pi S^{4}Q^{2}
	\nonumber \\ &\quad -\pi(a-1)l^{2}S^{4} +\pi^{2}(2a-5)l^{2}S^{3}Q^{2}.\nonumber
\end{align}
Comparing with $D_1$, we find $\mathrm{denom}(\tau)=-D_1$, so $\tau$ is singular where condition $II$ holds, $S=9.5986$ and $S=18.8026$, and the two stability criteria agree. In addition, $\tau$ vanishes at $S=1.2495$, which lies below the extremal entropy $S=1.9688$ and is therefore unphysical (see Fig.~\ref{f:3}).

\begin{figure*}[htbp]
	\centering
	\begin{subfigure}[b]{0.48\textwidth}
		\centering
		\includegraphics[width=\textwidth]{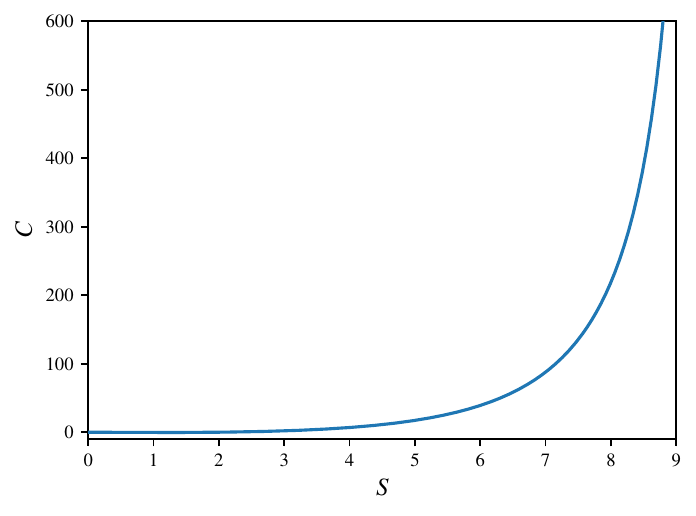}
		\caption{} 
		\label{f:4a}
	\end{subfigure}
	\hfill
	\begin{subfigure}[b]{0.48\textwidth}
		\centering
		\includegraphics[width=\textwidth]{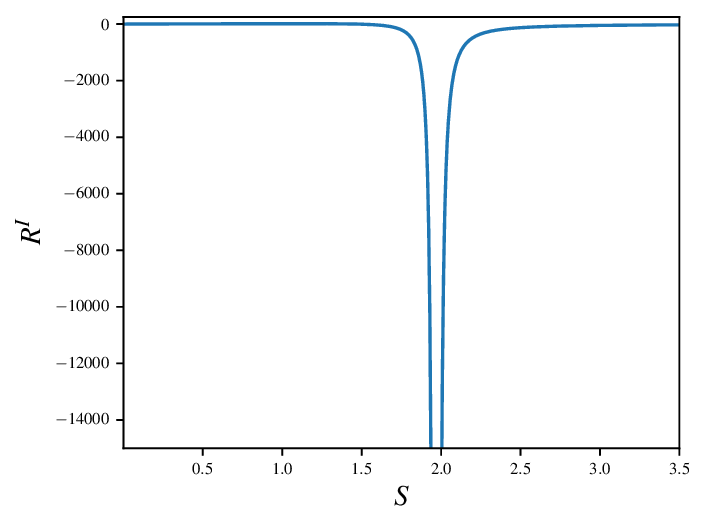}
		\caption{} 
		\label{f:4b}
	\end{subfigure}
	\caption{Behavior of (a) the heat capacity $C$ and (b) thermodynamic curvature $R^{\mathrm{I}}$ versus entropy $S$ for $Q=0.5$, $a=0.25$, and $l=6$, displaying $R^{\mathrm{I}}$ divergence coinciding with $C$ zeros.}
	\label{f:4}
\end{figure*}

We first consider the FOPT. According to Davies \cite{davies1977}, a FOPT takes place where the heat capacity vanishes together with the Hawking temperature. This is the case where condition $I$, Eq.~\eqref{e:21}, is satisfied: $T=C=0$ at the extremal point $S=1.9688$ (with $G=0.6361$), and only $S>1.9688$ is physical. In the GTD language, the divergences of the curvature scalar $R^{I}$ of the Quevedo-I metric $g^{I}$ therefore locate the FOPT, while condition $III$, Eq.~\eqref{e:23}, is not satisfied for any physical value of $S$, so $R^{III}$ gives no information. This gives a geometric meaning to the stability requirements of a system with three degrees of freedom \cite{Callen1985}, whose violation drives the system to a different thermodynamic state (Fig.~\ref{f:4}).

\begin{figure}[htbp]
	\centering
	\includegraphics[width=\columnwidth]{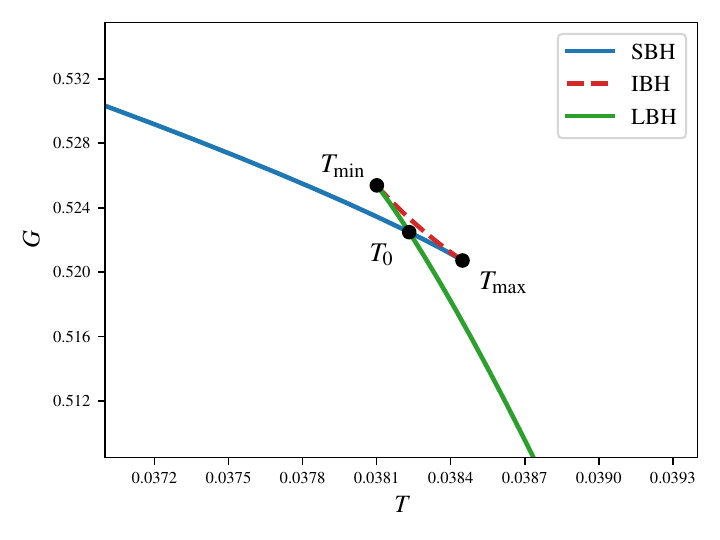}
	\caption{Behavior of the Gibbs free energy $G$ against temperature $T$ for $Q=0.5$, $a=0.25$, and $l=6$, displaying the thermodynamic phase structure.}
	\label{f:5}
\end{figure}

The FOPT can also be studied globally through the Gibbs free energy, Eq.~\eqref{e:26}, plotted against $T$ in Fig.~\ref{f:5}. Its swallow-tail shape signals a FOPT between a small and a large BH. The SBH and IBH branches meet at a cusp at $T_{\max}$, and the IBH and LBH branches at a cusp at $T_{\min}$ (Table~\ref{tab:1}). For $T_{\min}<T<T_{\max}$ all three branches coexist, but the IBH has the largest free energy and is never realized, so the SBH and LBH compete.

The free energies of the SBH and LBH are equal at the coexistence temperature $T_{0}=0.038232$, where the entropies are $S=7.5408$ (SBH) and $S=23.9124$ (LBH). Since $G$ is continuous at $T_0$ while the entropy jumps, this is a FOPT: the SBH is globally stable for $T<T_0$ and the LBH for $T>T_0$, each remaining metastable between $T_0$ and its own cusp. The window $T_{\max}-T_{\min}\approx 3.5\times10^{-4}$ is very narrow.

Comparing the BH phases with thermal AdS, whose free energy is taken as $G=0$, the LBH branch reaches $G=0$ at $T_{\rm HP}=0.046822$ ($S_{\rm HP}=92.145$), which lies outside the range of Fig.~\ref{f:5}. For $T>T_{\rm HP}$ the LBH dominates over thermal AdS, so a Hawking--Page-type transition \cite{Hawking1983} occurs at this point, at a higher temperature than the SBH/LBH transition at $T_0$, where $G>0$.

\begin{table}[htbp]
	\caption{Summary of characteristic points for $Q=0.5$, $a=0.25$, and $l=6$.}
	\label{tab:1}
	\begin{ruledtabular}
		\begin{tabular}{lccc}
			Point & $S$ & $T$ & $G$\\
			\hline
			Extremal ($T=0$) & $1.9688$ & $0$ & $0.6361$\\
			$T_{\max}$ & $9.5986$ & $0.038448$ & $0.5207$\\
			$T_{\min}$ & $18.8026$ & $0.038101$ & $0.5254$\\
			$T_{0}$ (SBH) & $7.5408$ & $0.038232$ & $0.5225$\\
			$T_{0}$ (LBH) & $23.9124$ & $0.038232$ & $0.5225$\\
			$T_{\rm HP}$ & $92.145$ & $0.046822$ & $0$\\
		\end{tabular}
	\end{ruledtabular}
\end{table}

\begin{figure*}[htbp]
	\centering
	\begin{subfigure}[b]{0.48\textwidth}
		\centering
		\includegraphics[width=\textwidth]{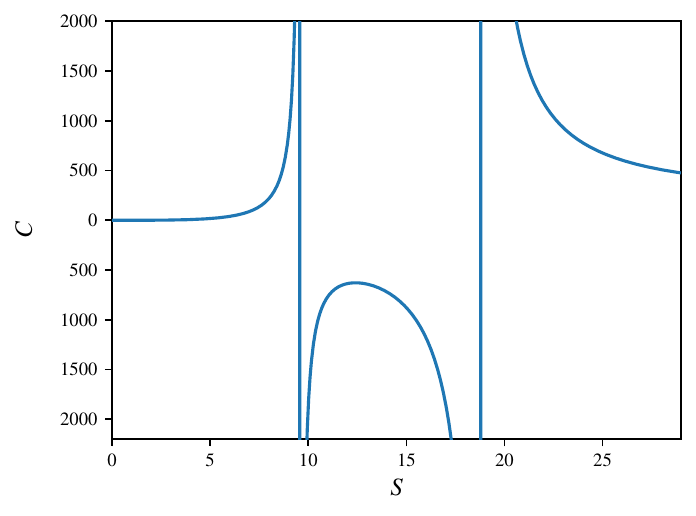}
		\caption{} 
		\label{f:6a}
	\end{subfigure}
	\hfill
	\begin{subfigure}[b]{0.48\textwidth}
		\centering
		\includegraphics[width=\textwidth]{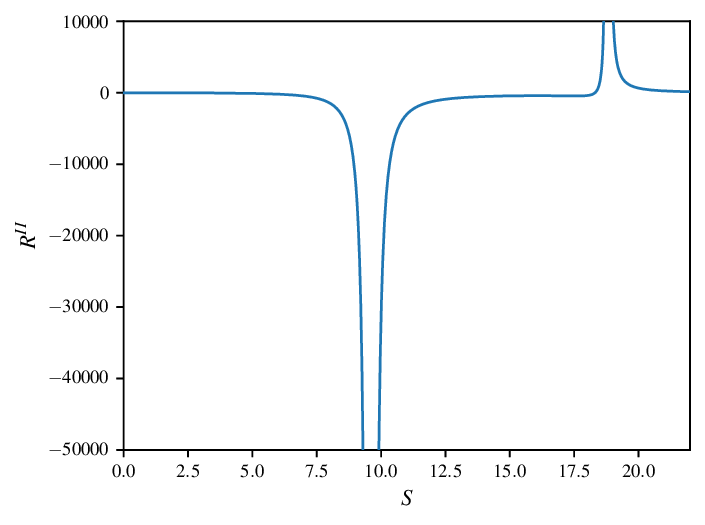}
		\caption{} 
		\label{f:6b}
	\end{subfigure}
	\caption{Behavior of (a) the heat capacity $C$ and (b) the thermodynamic curvature $R^{\mathrm{II}}$ versus entropy $S$ for $Q=0.5$, $a=0.25$, and $l=6$, illustrating the divergence points.}
	\label{f:6}
\end{figure*}

We now turn to the SOPT, which is detected through the response functions of the system. For a system with three degrees of freedom there are three independent response functions, and the divergence of the heat capacity is the standard signal of a SOPT. The heat capacity in Eq.~\eqref{e:25} becomes infinite whenever condition $II$, Eq.~\eqref{e:22}, is satisfied, which here happens at $S=9.5986$ and $S=18.8026$, the entropies of the cusps at $T_{\max}$ and $T_{\min}$. At both points $\partial T/\partial S=0$ and $C$ changes sign, so they mark the change of local stability and bound the metastable region of the FOPT. The same points are singular for the trace $\tau$ and coincide with the curvature singularities of the equilibrium manifold endowed with the metric $g^{II}$ (see Fig.~\ref{f:6}).

\begin{figure*}[htbp]
	\centering
	\begin{subfigure}[b]{0.48\textwidth}
		\centering
		\includegraphics[width=\textwidth]{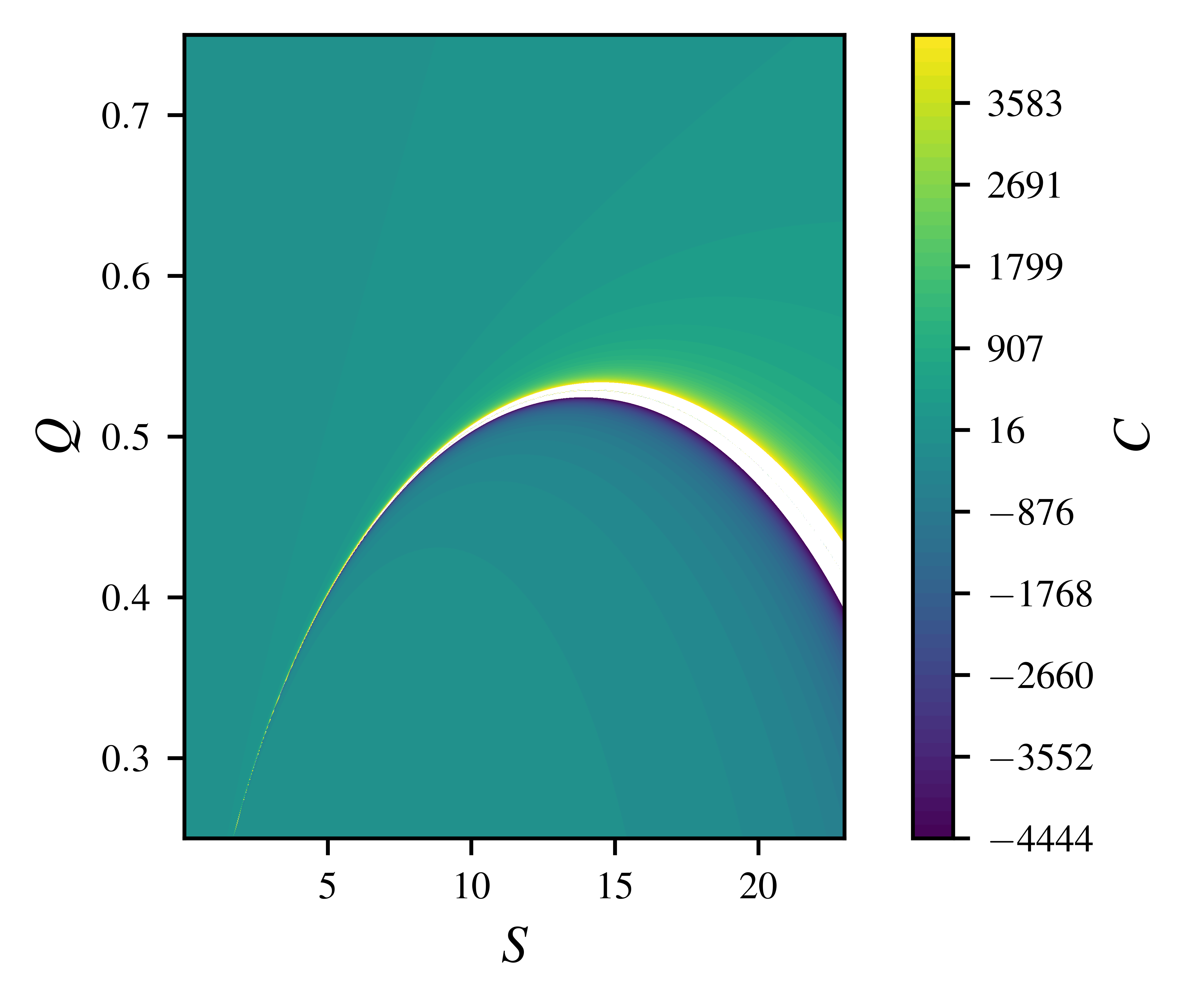}
		\caption{} 
		\label{f:7a}
	\end{subfigure}
	\hfill
	\begin{subfigure}[b]{0.48\textwidth}
		\centering
		\includegraphics[width=\textwidth]{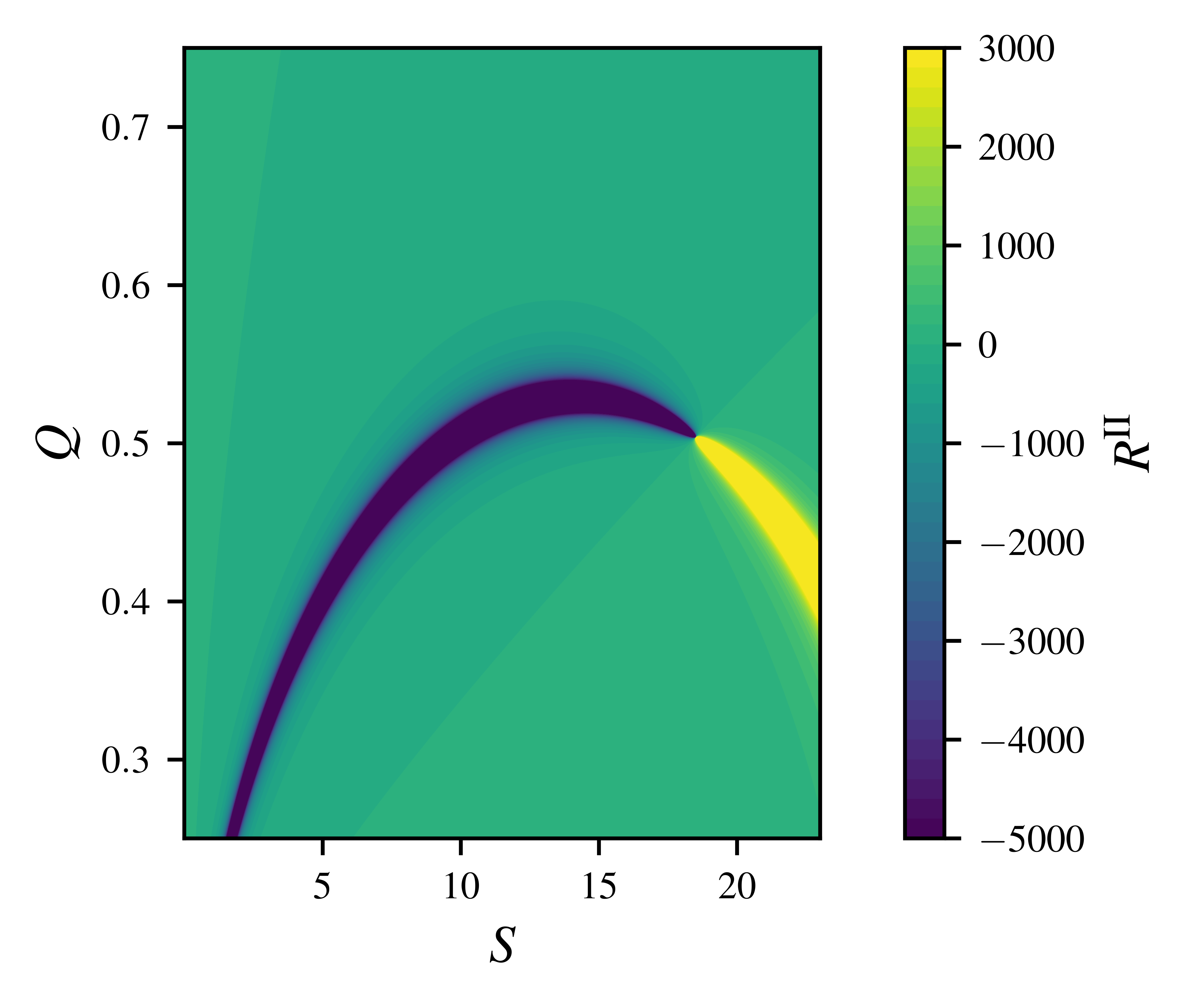}
		\caption{} 
		\label{f:7b}
	\end{subfigure}
	\caption{Contour profiles in the $S$--$Q$ plane displaying the variation of (a) the heat capacity $C$ and (b) the thermodynamic curvature $R^{\mathrm{II}}$ with fixed parameters $a=0.25$ and $l=6$.}
	\label{f:7}
\end{figure*}

The contour plots in Fig.~\ref{f:7} extend this comparison to the $S$--$Q$ plane. In Fig.~\ref{f:7a}, $C$ diverges along an arch-shaped curve that separates the region of positive $C$ (outside) from that of negative $C$ (inside), and its two legs are the roots of $D_1$, i.e.\ the SOPT points at $S=9.5986$ and $S=18.8026$ for $Q=0.5$. The singularities of $R^{II}$ in Fig.~\ref{f:7b} follow the same arch. The two roots merge at the apex, $Q_c\approx0.53$ and $S\approx14$, so for $Q>Q_c$ neither the divergences nor the swallow tail exist. Since the Quevedo metrics are Legendre invariant, this agreement does not depend on the choice of thermodynamic potential.

\section{Conclusions} \label{sec:5}

We have studied how a surrounding cloud of strings reshapes the thermodynamics of the ABG black hole in anti-de Sitter space, and in particular its phase transitions, using GTD. Rewriting the metric function in a suitable form allowed us to take the black hole mass $M$ as the fundamental relation of the system. This mass is a quasi-homogeneous function, so a single expression contains the entire thermodynamic content of the configuration. Because the AdS radius $l$ enters every thermodynamic quantity in a significant way, we included it among the independent variables of the thermodynamic phase space instead of keeping it fixed.

We then constructed the equilibrium manifold, equipped it with each of the three GTD metrics, and computed the corresponding scalar curvatures, whose divergences signal a change of phase or of thermodynamic stability. The $g^I$ and $g^{II}$ metrics gave two distinct nontrivial divergence conditions, whereas the condition of the $g^{III}$ metric is not satisfied for any physical value of the entropy. For $Q=0.5$, $a=0.25$, and $l=6$, $R^{I}$ diverges at the extremal point, where $T$ and $C$ vanish together, in agreement with the FOPT condition of Davies, and $R^{II}$ diverges exactly where $C$ and the trace of the Hessian matrix of the Gibbs free energy diverge, which locates the SOPT points. The contour plots show that this agreement is not restricted to a single choice of parameters.

The Gibbs free energy completes the picture. Its swallow-tail structure shows that the small and large black holes exchange global stability at $T_{0}=0.038232$, which is a FOPT, and that the divergences of $C$ bound the metastable region of this transition. At the higher temperature $T_{\rm HP}=0.046822$ a Hawking--Page-type transition from thermal AdS to the large black hole takes place.

Overall, once the fundamental equation is known, GTD can describe the phase structure of a black hole purely through geometry, even in the presence of additional matter such as a cloud of strings. It would be worthwhile to study the quantum-corrected ABG--AdS black hole, including higher-order curvature corrections of the Gauss--Bonnet type, and to apply the quasi-homogeneous version of GTD to other black hole solutions of a similar kind.
\section*{Funding Declaration}
The authors received no external funding for this work.
\section*{Declaration of competing interest}
The authors declare no competing financial or personal interests that could have influenced this work.
\section*{Data availability}
No data was used for the research described in the article.

\bibliographystyle{apsrev4-2}
\bibliography{bibliography}
\end{document}